\documentclass[amsmath,amssymb,aps,prl,twocolumn, superscriptaddress,longbibliography]{revtex4-2}
\usepackage{mathrsfs}
\usepackage{graphicx}
\usepackage{dcolumn}
\usepackage{color}
\usepackage{hyperref}
\usepackage{enumitem}

\newcommand{\<}{\langle}
\renewcommand{\>}{\rangle}

\renewcommand{\cite}[1]{[\onlinecite{#1}]}
\newcommand{\cra}[2]{\hat{#1}^{\dag}_{#2}}  
\newcommand{\ana}[2]{\hat{#1}_{#2}}         
\newcommand{\bra}[1]{\left<#1\right|}      
\newcommand{\ket}[1]{\left|#1\right>}      

\newcommand{\eps}{\varepsilon}      
\newcommand{\sgn}{{\rm {sgn}}}

\begin{document}

\title{Vortex bound states in dimerized $\pi$-flux optical lattices: \\ characterization, state preparation and current measurement}

\author{Andrei A. Stepanenko}
\affiliation{London Institute for Mathematical Sciences, Royal Institution, 21 Albemarle St, London W1S 4BS, UK}
\affiliation{School of Physics and Engineering, ITMO University, Saint Petersburg 197101, Russia}

\author{Marco Di Liberto}
\affiliation{Dipartimento di Fisica e Astronomia “G. Galilei” \& Padua Quantum Technologies Research Center, Università degli Studi di Padova, I-35131, Padova, Italy}
\affiliation{Istituto Nazionale di Fisica Nucleare (INFN), Sezione di Padova, I-35131 Padova, Italy}

\date{\today}

\begin{abstract}
Lattice models display bound states for repulsive interactions that smoothly connect to high-energy two-particle states of doubly occupied sites, namely doublons, for strong onsite interactions. 
In this work, we show that a distinct type of repulsively bound states, namely vortex bound states of two bosons, appear in dimerized square lattices pierced by a uniform $\pi$-flux for moderate interactions. 
By focusing on a ladder geometry as an illustrative example, we characterize their properties, including chirality-changing decay channels induced by flux detuning, and we develop protocols to perform state preparation in optical lattices via adiabatic sequences or recently developed current imprinting methods. 
Finally, we show how to measure currents and thus chirality by quenching the system onto isolated pairs of nearest-neighbor sites and then sampling the corresponding dynamics.
These results can also provide an experimentally realistic strategy for state preparation and probing of chiral gapped many-body phases in optical lattices. 
\end{abstract}

\maketitle

\emph{Introduction.} Quantum simulation architectures have shown remarkable success in accessing and controlling many-body quantum phenomena \cite{Georgescu2014Mar, Altman2021Feb}.
Among these, atom-based platforms like ultracold atoms in optical lattices \cite{Gross2017} are a versatile setting to study ground state and out-of-equilibrium properties of Hubbard models and their generalizations \cite{Dutta2015}.
A central goal of quantum simulation with ultracold atoms has been to investigate quantum states that break time-reversal symmetry, including the role played by interactions \cite{Goldman2016,Cooper2019}. 
For instance, weak repulsive interactions in higher bands favour the formation of superfluid phases of vortex arrays \cite{Wirth2011,Li2016,DiLiberto2016,Xu2021}.
However, strongly-interacting states with broken time-reversal symmetry \cite{Rachel2018}, like chiral Mott insulators \cite{Dhar2012}, vortex and Meissner phases \cite{Petrescu2013, Piraud2015, Celi2023} or quantum Hall phases \cite{Raghu2008} remain elusive.
Thanks to the advances in the simulation of synthetic gauge fields~\cite{Goldman2014}, recent developments in this direction have made possible to prepare Laughlin states of few bosonic particles~\cite{Leonard2023}. 
Fermionic ones have instead being obtained in optical tweezers by imprinting angular momentum via Laguerre-Gaussian beams~\cite{Jochim2024}.
These represent first steps towards the simulation of fractional Quantum Hall phases \cite{Laughlin1983}.

\begin{figure}[t]
\includegraphics[width = \columnwidth]{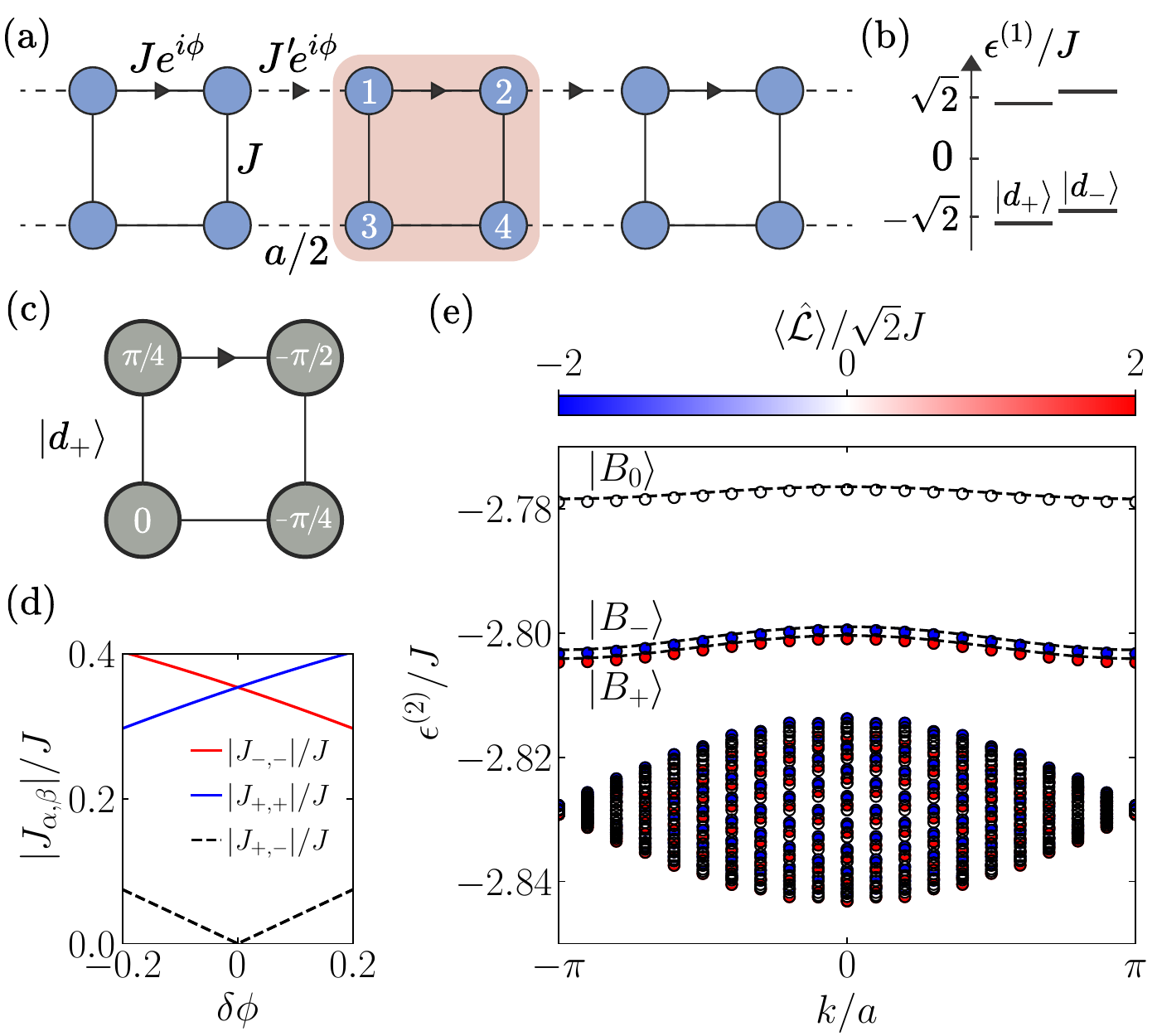}
\caption{\label{fig:model} 
Model. (a) Ladder geometry. 
(b) Single-particle spectrum of a four-site plaquette with flux $\phi=\pi+\delta\phi$.
(c) Phase structure of a single-particle chiral state $|d_+\rangle$.
(d) Effective theory couplings with a finite flux detuning $\delta\phi$.
(e) Low-energy two-body spectrum for $J'/J = 0.01$, $U/J =0.1$ and a small flux detuning $\delta \phi =10^{-3}$ for $N=20$ plaquettes.
Dashed lines are analytical results obtained from the projected theory.
}
\end{figure}

Few particle systems provide a useful testbed to study intriguing interacting phenomena, as for the case of repulsively lattice bound states \cite{Mattis1986,Winkler2006,Valiente2008}.
Due to the short-range nature of interactions in alkali atoms \cite{Bloch2008}, these bound states typically occur in the form of long-lived doublons, which behave as simple composite particles with a renormalized mass or hopping. 
These features reveal themselves in the two-particle quantum walk dynamics or via the presence of fractional Bloch oscillations \cite{Corrielli2013,Greiner2015}.
However, there are many cases where the internal structure of the bound states, manifesting in a finite probability of finding the two particles at a finite distance, really matters.
In these situations, remarkable differences with a simple composite particle picture emerge.
For example, bound states in the presence of artificial magnetic fields can manifest different chiral properties or change topological sector \cite{Bello2017, Tai2017, Lee2017, Salerno2018} as compared to the noninteracting case. 
In dimerized chains, expected topological edge bound states are instead unprotected and resonant with the bulk modes \cite{Bello2016, DiLiberto2016b, Gorlach2017}.
Strong nearest-neighbor interactions make two particles mainly bind at a finite relative distance \cite{Valiente2009,DiLiberto2017,weckesser2024}, dramatically altering the effective lattice geometry and inducing localization or emerging topological properties \cite{Salerno2020}. 
These can also occur via the interplay of different interaction channels \cite{Stepanenko2020, Stepanenko2022}.
All these considerations raise the question of identifying under which conditions bound pairs can have an internal structure that breaks time-reversal symmetry thus displaying a finite angular momentum.

In this work, we positively address this question and show that \emph{vortex bound states} of two bosonic particles exist in lattice models with dimerized couplings and uniform $\pi$-flux.
We further show resonances with scattering states and pair conversion induced by small flux detunings.
We discuss state preparation of vortex bound states via two complementary protocols: an adiabatic sequence and a current imprinting strategy recently demonstrated in experiments. 
Finally, we develop a convenient current measurement scheme that is applicable to fully interacting systems and in the absence of parity projection \cite{Atala2014Aug,Impertro2023Dec}.
Our results offer a distinct direction towards the realization and control of interacting chiral states of matter, especially in view of long-sought realization of interacting chiral gapped phases \cite{Piraud2015,DiLiberto2023}.

\emph{Model.} We study the dimerized ladder model with uniform flux shown in Fig.~\ref{fig:model}(a), inspired by the BBH model introduced in the context of higher-order topological insulators~\cite{Benalcazar2017Jul}. 
We consider two interacting bosonic particles with repulsive Hubbard interaction $\hat H_\text{int}=(U/2)\sum_{i} \hat{n}_{i}(\hat{n}_{i}-1)$.
We focus on the regime near $\pi$-flux, namely $\delta \phi \equiv \phi-\pi \ll 1$, and strong dimerization $J'\ll J$, such that for each plaquette $m$ with strong hopping $J$, the four single-particle eigenstates are grouped in pairs and are separated by a gap $\Delta E_\text{gap}\approx 2\sqrt{2}J$ (see Fig.~\ref{fig:model}(b)). 
Similarly to the discussion in Ref.~\cite{DiLiberto2023}, the energy-scale separation provided by the gap allows us to consider the lowest two eigenmodes $|d_{m,\pm} \rangle$ to construct a projected low-energy theory when $J'$ and $U$ are small as compared to $\Delta E_\text{gap}$.

The single particle modes display the real-space structure $|d_{m,\pm} \rangle = \sum_{\sigma=1}^4 \alpha_{\pm,\sigma} |b_{m,\sigma} \rangle$, see Fig.~\ref{fig:model}(c), where $|b_{m\sigma}\rangle$ indicates a single-particle boson in the sublattice $\sigma$ of the $m$th-plaquette, $\alpha_\pm = (u_\pm,v_\pm,1,u_\pm^*)/2$ with $u_\pm = e^{i(\pi\pm\pi-\phi)/4}$ and $v_\pm = e^{i(\pi\pm\pi+\phi)/2}$. 
For $\phi=\pi$, the two eigenstates are degenerate in energy and are time-reversal partners. 
Each mode breaks time-reversal symmetry and thus carries the notion of chirality. 
We quantify this physical property in real space by defining the plaquette bond currents
$\hat j^{(m)}_{\sigma,\sigma'} = i (J_{\sigma,\sigma'} \hat b^\dagger_{m,\sigma} \hat b^{}_{m,\sigma'}- \text{H.c.})$, which we translate into a loop-current operator $\mathcal{\hat L}_m = \sum_{\sigma,\sigma'}' \hat j_{\sigma,\sigma'}^{(m)}$, where $\sum'$ indicates a clockwise loop. 
Within the projected theory and for $\phi=\pi$, one can show that the loop current is related to the relative occupation of the two chiral modes, i.e. the chiral population difference $\hat{L}_{z,m} \equiv \cra{d}{m,+}\ana{d}{m,+} - \cra{d}{m,-}\ana{d}{m,-}$, via the relation $\hat{\mathcal{L}}_m \approx \sqrt{2}J\hat{L}_{z,m}$. 
In this sense, the two modes carry a unit of loop current with opposite signs and play a similar role as the $p_\pm = p_x\pm i p_y$ orbitals in $p$-band models~\cite{Wu2015Jun,Sala2015Mar,DiLiberto2023}. 
For this reason, we will refer to $\hat{L}_{z,m}$ as the \emph{angular momentum} operator.
However, the proportionality between the loop current operator and the angular momentum operator is not valid for $\phi\neq\pi$ or beyond the validity of the effective theory, and deviations take place, as we will show below.

Under the assumption $\delta \phi\ll 1$ and $J', U \ll J$, a projected low-energy theory takes the form \cite{DiLiberto2023}
\begin{eqnarray}
   \hat{H}_\text{eff}&=&  
   \sum_{m,\alpha, \beta} \left( J_{\alpha,\beta}(\phi) \cra{d}{m+1,\alpha}\ana{d}{m,\beta} +\mathrm{H.c.} \right) \nonumber \\&& 
   +\sum_{m,\alpha} \mu_{\alpha}(\phi) \cra{d}{m,\alpha}\ana{d}{m,\alpha}\nonumber\\&&
    + U\sum_m\left(\dfrac{3\hat{n}_m^2}{16} - \dfrac{\hat{L}_{z,m}^2}{16}  - \dfrac{\hat{n}_m}{8} \right)\,,    
   \label{eq:Ham_proj}
\end{eqnarray}
where $\alpha,\beta\in\left\{+,-\right\}$ label the two lowest eigenmodes and we introduced the projected particle number operator $\hat{n}_m = \cra{d}{m,+}\ana{d}{m,+} + \cra{d}{m,-}\ana{d}{m,-}$.
The specific dependence of the parameters $J_{\alpha,\beta}(\phi)$ and $\mu(\phi)$ is given in Ref.~\cite{SuppMat}.
In the limit $\phi\rightarrow{\pi}$ the orbitals decouple, namely $J_{\pm,\mp}\rightarrow 0 $, and $|J_{-,-}| = |J_{+,+}|$ as in Ref.~\cite{DiLiberto2023}. 
The dependence of the hopping coefficients on the flux $\phi$ is shown in Fig.~\ref{fig:model}(d).
The last line describing two-particle interactions shows the appearance of a new term as compared to typical Bose-Hubbard models, namely an angular momentum term analogous to the one appearing in $p$-band models \cite{Liu2006}, which will determine the bound states structure.

\emph{Two-particle states.} The Bose-Hubbard model is known to feature repulsively bound states in its spectrum~\cite{Winkler2006}. 
When onsite interactions dominate over hopping, these states are described by mobile doublons.
Here, however, the dominant energy scale is $J$ and we thus expect bound states delocalized over the entire four sites of each plaquette, similarly to what occurs in the SSH model case for its two-sites unit cell~\cite{DiLiberto2016b, Gorlach2017}.
The peculiar angular momentum dependence of the interaction channel will favour bound states that can display a non-vanishing chirality, namely \emph{vortex bound states}.

To clarify this mechanism, let us consider the $J'=0$ limit first.
Since in each plaquette $m$ we can define single-particle eigenstates possessing angular momentum $\langle \hat L_{z,m}\rangle = \pm 1$, three kinds of two-particle states are possible. 
Two states, $\ket{B_{m,\pm}} = (\cra{d}{m,\pm})^2|0\rangle/\sqrt{2}$, display finite and opposite angular momentum (or chirality) with $\langle \hat L_{z,m}\rangle = \pm 2$. A third state,  $\ket{B_{m,0}} = \cra{d}{m,+}\cra{d}{m,-}|0\rangle$, instead has no angular momentum $\langle \hat L_{z,m}\rangle = 0$.
Since interactions nonlinearly depend on the angular momentum, the two chiral bound states $\ket{B_{m,\pm}}$ are eigenstates with energies $\epsilon_\pm^{(2)} = 2\mu_\pm+U/4$, whereas the non-chiral one $\ket{B_{m,0}}$ has a distinct energy $\epsilon_0^{(2)} = \mu_++\mu_-+U/2$.
When $J'\neq 0$, the two-particle problem can be solved exactly within the projected theory starting from the Schr\"odinger equation $\hat{H}_\text{eff}\ket{\psi} = \epsilon^{(2)}\ket{\psi}$ and searching for a two-particle wavefunction solution of the form 
$\ket{\psi} = \sum_{\alpha,\beta,m, n}C_{\alpha\beta}e^{ik(m+n)/2}e^{i\kappa(n-m)/2}\cra{d}{m,\alpha}\cra{d}{n,\beta}\ket{0}$, where $k$ and $\kappa$ parametrize the center-of-mass momentum and the relative momentum, respectively.
In Fig.~\ref{fig:model}(e), we show the three bound states energy dispersions $\epsilon_\mu^{(2)}(k)$, with $\mu=0,\pm$, more details are provided in Ref.~\cite{SuppMat}.
 
In addition to bound states, the system displays two-particle scattering or quasi-independent states. 
Scattering states can be classified according to their chirality in a similar fashion as for bound states, and we indicate those as $\ket{S_\pm}$ and $\ket{S_0}$. 
The low-energy spectrum of the two-particle problem discussed so far is shown in Fig.~\ref{fig:model}(e) for an array of 20 plaquettes. 
According to the projected  model~\eqref{eq:Ham_proj}, where the onsite energy of each mode depends on the flux through the quantity $\mu_\alpha(\phi)$, the energies of the chiral states $\ket{B_\pm}$ and $\ket{S_\pm}$ depend on the flux, but the energies of the states $\ket{B_0}$ and $\ket{S_0}$ remain unaffected. 
This opens intriguing possibilities concerning the interplay between interactions, angular momentum and flux. 
The full spectrum is quite reach and we comment on the high energy states in Ref.~\cite{SuppMat}. 

\begin{figure}[t]
\includegraphics[width = \columnwidth]{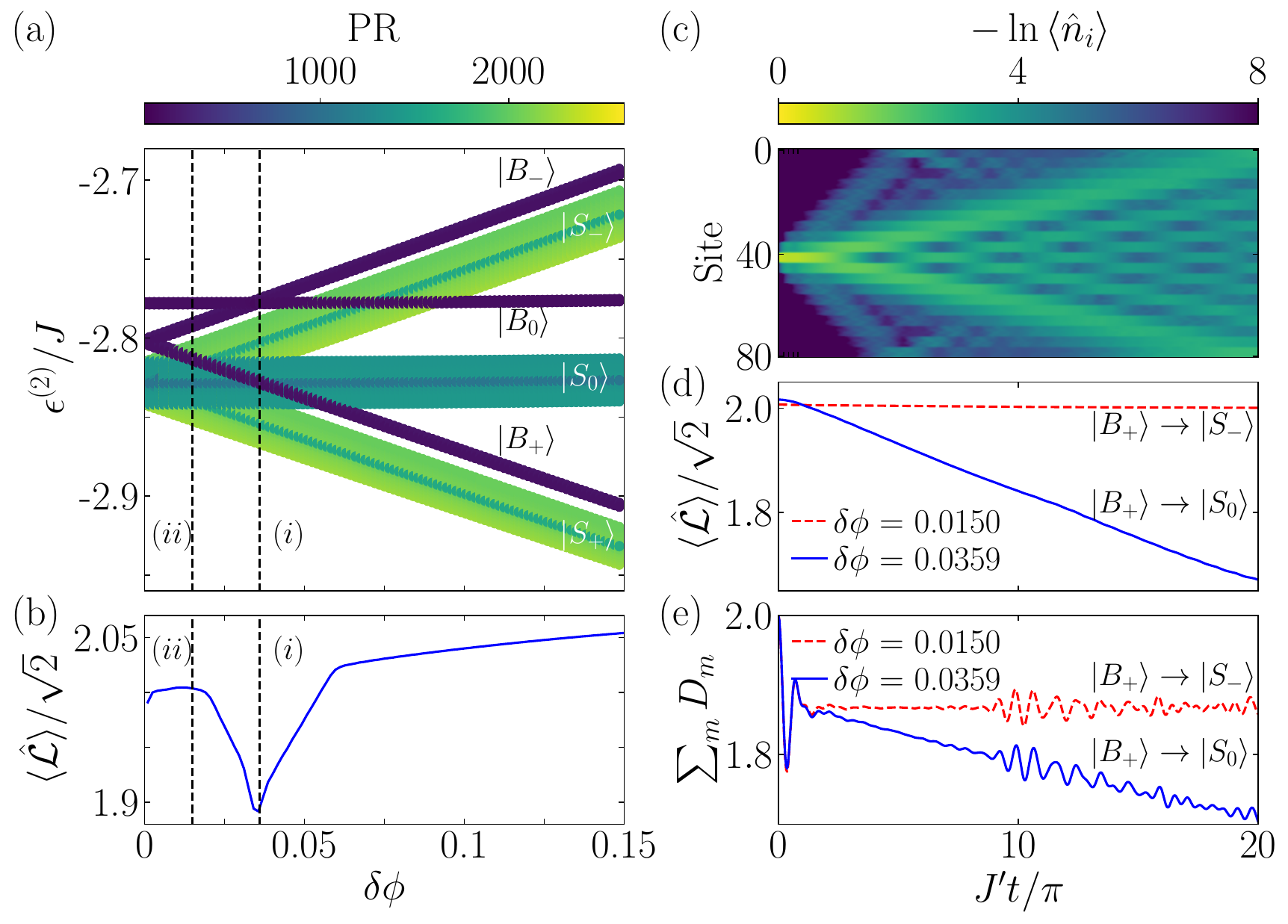}
\caption{\label{fig:3} 
Finite flux detuning effects.
(a) Two body spectrum at finite flux for  $J'/J = 0.01$, $U/J = 0.1$ and $N=20$ plaquettes (open boundary conditions) .
Color represents the participation ratio $\text{PR}\equiv(\sum_m(|\psi_m|^4))^{-1}$ showing extended and bound states.
Dashed lines indicate the two resonances discussed in the text.
(b) Loop current at time $t= 7\pi/J'$ of an initially prepared $|B_{m,+}\rangle$ state showing resonant conversion to $S_0$ scattering states in a range of flux detuning values.
(c) Time evolution of the particle density for the $B_+$ state with $\delta\phi = 0.0359$ .
(d) Time evolution of the loop current operator and of (e) the density-density correlator for the two values of flux indicated by $(i)$ and $(ii)$ in (a) and (b).
}
\end{figure}

\emph{Resonances from flux detuning.} For finite values of $\delta \phi\neq 0$, the non-vanishing coefficients $J_{+,-}$ in the model \eqref{eq:Ham_proj} break the angular momentum conservation.
Under a positive ($\delta\phi >0$) change of flux, several distinct regimes can thus occur depending on different energy crossings. 
Among all the possibilities, we identify and analyse two resonances of interest: a chiral bound state band (e.g. $\ket{B_+}$) crossing $(i)$ a continuum of non-chiral scattering states ($\ket{S_0}$) or $(ii)$ a continuum of chiral scattering states with opposite chirality ($\ket{S_-}$).
From Fig.~\ref{fig:3}(a), we see that two other possible types of resonances can occur.
The first of these takes place between a non-chiral bound state ($\ket{B_0}$) and a continuum of chiral scattering states ($\ket{S_-}$).
The second one is a resonance between two bound states ($\ket{B_0}$ and $\ket{B_-}$).
In the following, we focus on cases $(i)$ and $(ii)$.

We test the two regimes presented above by preparing the initial state $(\cra{d}{m,+})^2|0\rangle/\sqrt{2}$ for a plaquette $m$ in the bulk of the ladder, which has a large overlap with the bound state manifold $\ket{B_+}$ when $U \gg J'$.
We find that the conversion process $(i)$ $\ket{B_+}\rightarrow \ket{S_0}$ displays a resonant drop of loop current $\langle \hat{\mathcal{L}} \rangle = \sum_m \langle \hat{\mathcal{L}}_mw \rangle$ (see Fig.~\ref{fig:3}(b)) at the value of $\delta \phi$ corresponding to energy crossing.
Instead, the process $(ii)$ $\ket{B_+}\rightarrow \ket{S_-}$ does not take place and the angular momentum is conserved over time, as such process would require the chirality flip of both particles and is perturbatively suppressed.
In Fig.~\ref{fig:3}(c), an example of two-particle quantum walk dynamics is shown.
The loop current, shown in Fig.~\ref{fig:3}(d), linearly drops over time for the resonant parameter choice, case $(i)$, while it remains constantly stable far away from the resonant condition.
The plaquette correlator $
    D_m \equiv \frac 1 2 \sum_{\sigma,\sigma'} \langle \hat{n}_{m,\sigma}\hat{n}_{m,\sigma'} \rangle
$, which should be large for bound states, and drop for quasi-independent particles, confirms this conclusion as shown in Fig.~\ref{fig:3}(e).

\emph{State preparation.} In order to prepare a vortex bound state, the simplest strategy is to follow and adiabatic ramp in a single disconnected plaquette with initial finite flux detuning. 
The details are discussed in the End Matter section.
A distinct approach can be pursued by a digital protocol imprinting particle currents, which builds on the recent experiment from Ref.~\cite{Impertro2023Dec} that we briefly review. 
Consider two sites $a$ and $b$, forming a two-level system or \emph{qubit}, of which we can separately control the hopping and the onsite energy difference via the two unitary operations
\begin{eqnarray}
    \hat{\mathcal U}_{ab}^X(t)&\equiv& \exp\left(-i (-J_{ab}\cra{b}{a}\ana{b}{b}^{}+\text{h.c.})t/\hbar\right)\,,\\
    \hat{\mathcal U}_{ab}^Z(t)&\equiv& \exp\left(-i \Delta (\hat{n}_a - \hat{n}_b)t/2\hbar\right)\,.
\end{eqnarray}
If we start with a particle localized on site $a$, $|\psi(0)\rangle = |a\rangle$, we can prepare a uniform two-site state, namely an equatorial state of the qubit, via a pulse
$\hat{\mathcal U}_{ab}^X(Jt=\pi/4) |a\rangle = \frac{1}{\sqrt 2}\left(|a\rangle+e^{i\pi/2}|b\rangle\right)$,
where we took for simplicity $J_{ab}=J$ and $\hbar=1$.
As this corresponds to a $\pi/2$ rotation on the Bloch sphere around the $x$-axis, we will thus indicate it as $\hat X^{ab}_{\pi/2}\equiv \hat{\mathcal U}_{ab}^X(Jt\!=\!\pi/4)$.
We can adjust the relative phase (and thus the value of the current $\langle \hat j_{ab} \rangle$) by subsequently applying $\hat{\mathcal U}_{ab}^Z(t)$ for the required time.
In this case, a rotation of an angle $\theta$ around the $z$-axis is denoted as $\hat Z^{ab}_\theta\equiv \hat{\mathcal U}_{ab}^Z(\Delta t\!=\!\theta)$.
\begin{figure}[!t]
\includegraphics[width = \columnwidth]{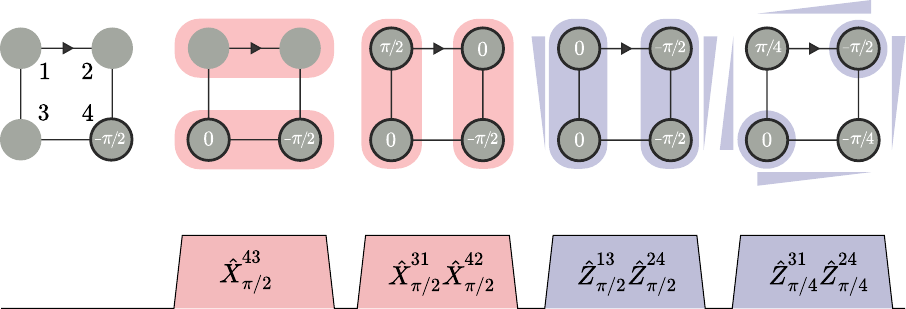}
\caption{
\label{fig:5}
Digital preparation.
Current imprint protocol for state preparation of the single-particle target state $|d_+\rangle$ shown in Fig.~\ref{fig:model}(c). 
Occupied sites with the same density are represented by black circles.
The initial state consists of a particle on site 4 (black circle). 
Four pulses according to the definition used in the text and involving pairs of sites transport the particle via hopping processes (red shadows) and change the phases (violet shadows) via onsite energy gradients represented by triangles. 
The final state is the one in Fig.~\ref{fig:model}(c).
}
\end{figure}
Our target is a uniform state on all four sites of a plaquette with a phase pattern given by one of the chiral states, e.g. $|d_+\rangle \sim e^{i\pi/4}|b_1\rangle + e^{i3\pi/2}|b_2\rangle + |b_3\rangle + e^{-i\pi/4}|b_4\rangle$, shown in Fig.~\ref{fig:model}(c).
This can be achieved by the following steps:
\begin{enumerate}[label=\Roman*.]
    \item A particle, initially localized on site 4, is transported into an equal superposition of sites 3 and 4 via a pulse 
    $\hat X^{43}_{\pi/2}$;
    \item A coherent superposition with sites $1,\,2$ is obtained by a combined pulse 
    $\hat X^{31}_{\pi/2} \hat X^{42}_{\pi/2}$;
    \item The phases of the newly populated sites are shifted by 
    $\hat Z^{13}_{\pi/2}\hat Z^{24}_{\pi/2}$;
    \item The phases of the \emph{diagonal sites} $1$ and $4$ are shifted by 
    $\hat Z^{31}_{\pi/4}\hat Z^{24}_{\pi/4}$.
\end{enumerate}
Adiabatically ramping up the interactions via a Feshabch resonance at the end of the sequence allows to connect to the interacting vortex bound state.
The protocol discussed above, and summarized in Fig.~\ref{fig:5}, is specifically designed for experimental setups where one can simultaneously operate on all lattice sites \emph{pairwise}, namely on pairs of dimer sites. 
This can be realized efficiently using superlattices, where tunnel couplings and onsite energies can be applied homogeneously for large systems in parallel on all dimers~\cite{Atala2014Aug,Impertro2023Dec}. 
Step IV however requires projecting local potentials into the atomic plane.

\begin{figure}[!b]
\includegraphics[width = \columnwidth]{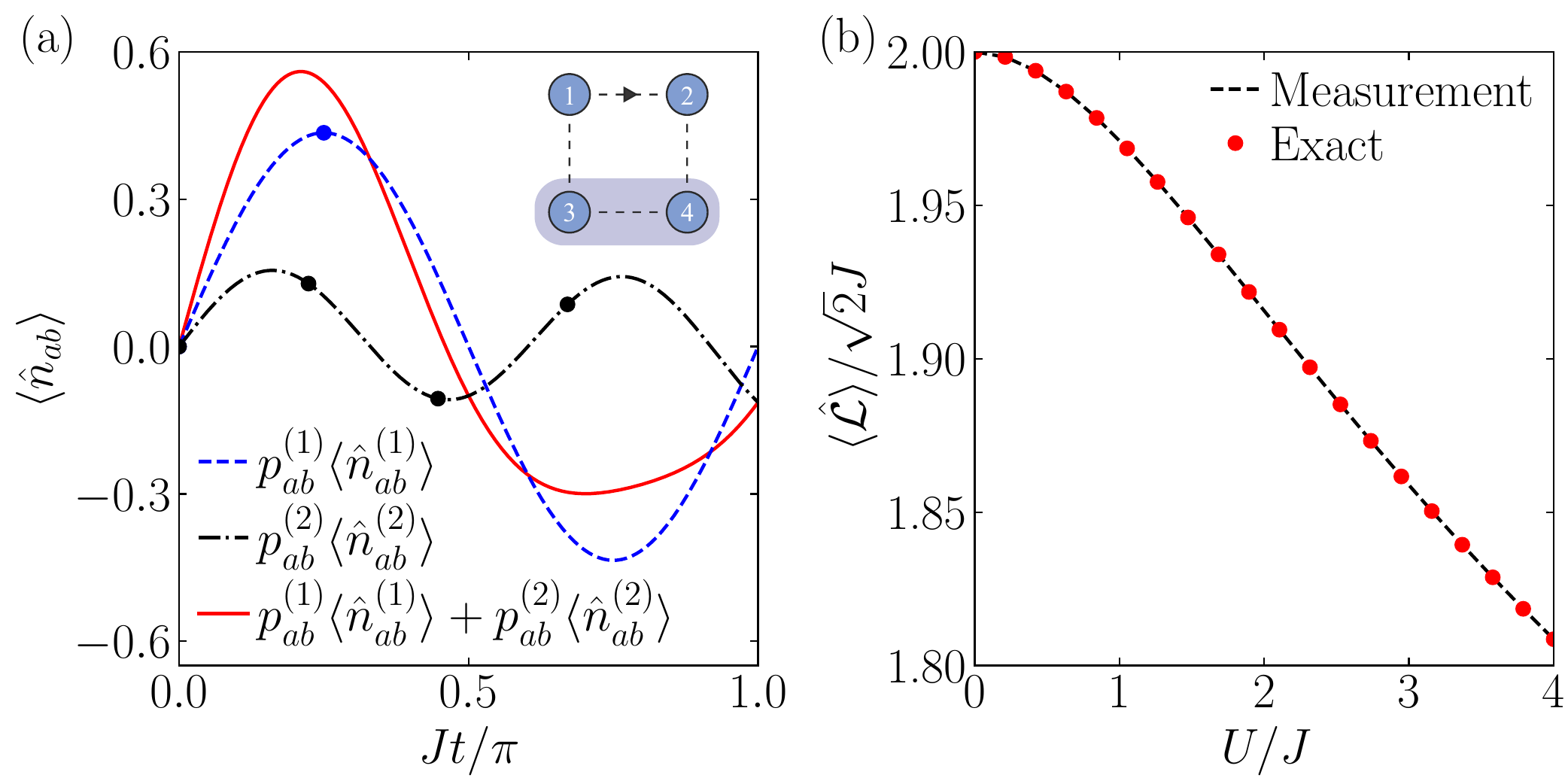}
\caption{\label{fig:meas} 
(a) Dynamics of the differential density for two nearest-neighbor sites within a plaquette (see inset).
The different curves show the one-particle sector (dashed line), the two-particle sector (dash-dotted line) and the total differential density (solid red line). 
(b) Comparison between the plaquette loop current $\<\hat{\mathcal{L}}\>$ reconstructed from differential density measurements (dashed black line) and the theoretical value obtained via exact diagonalization (red dots) for increasing interactions strengths.}
\end{figure}
\emph{Chirality measurement.} Probing the chiral property of a quantum state can be achieved by measuring the local bond currents. 
Reducing the occupation to a maximum of two particles, the current can be decomposed into 
\begin{equation}
    \langle \hat{j}_{ab} \rangle (\tau_0) = p_{ab}^{(1)} \mathcal{J}_0^{(1)} + p_{ab}^{(2)} \mathcal{J}_{0}^{(2)} \,,
\end{equation}
where $p_{ab}^{(1)}$ and $p_{ab}^{(2)}$ are the probability to find one or two particles in the $a$, $b$ sites, respectively.
In Ref.~\cite{Atala2014Aug} it was shown that the single-particle contribution $\mathcal{J}_0^{(1)}$ can be obtained by quenching the system into isolated dimers and measuring the oscillating density.
More specifically, one obtains $\mathcal{J}_0^{(1)} = J\mathcal{N}_1^{(1)}$, where $\mathcal{N}_1^{(1)}\equiv\langle\hat{n}_{a}-\hat{n}_{b}\rangle(\tau_1)$ is the differential density measured at time $\tau_1 = \pi/4J$.
This works also for the many-particle case, provided that interactions are turned off via a Feshbach resonance ($U\rightarrow0$) when the quench dynamics takes place.
This procedure, which slows down the experimental sequence, can be avoided by including the interacting contribution $\mathcal{J}_{0}^{(2)}$ provided that parity projection is not in place in the quantum gas microscope~\cite{gross_quantum_2021}. 
By exploiting the continuity equation, we find this contribution to be 
\begin{align} 
    \mathcal{J}_{0}^{(2)} 
    =&-\csc{\left(\frac{U\pi}{2\omega}\right)}\left[
    \dfrac{U}{4}\mathcal{N}_{0}^{(2)}
    \cos{\left(\frac{U\pi}{2\omega}\right)}\right.
    +\dfrac{U}{4}\mathcal{N}_{2}^{(2)}\nonumber\\&
    -\dfrac{\omega}{2} \mathcal{N}_{1}^{(2)}\sin{\left(\frac{3U\pi}{4\omega}\right)}
    \left.-\dfrac{\omega}{2} \mathcal{N}_{3}^{(2)}\sin{\left(\frac{U\pi}{4\omega}\right)}
    \right]\,,
    \label{eq:j2p0}
\end{align}
showing that it is enough to measure the differential density, $\mathcal{N}_{n}^{(2)}$, at four times $\tau_n$, where $\tau_n\equiv n \pi/2\omega$, with $\omega \equiv \sqrt{U^2+16J^2}$ and $n=0,\dots, 3$.
In Fig.~\ref{fig:meas}, we apply this protocol to a $\pi$-flux plaquette where the two-particle state $|B_+\rangle$ is initially prepared. 
In Fig.~\ref{fig:meas}(a), we show the differential density time evolution on a pair of sites for the single-particle sector, $\langle\hat{n}_{ab}^{(1)}\rangle$, and the two-particle sector, $\langle\hat{n}_{ab}^{(2)}\rangle$, respectively. 
In Fig.~\ref{fig:meas}(b), we compare the  the measurement protocol with the exact results and show the dependence of the plaquette loop current $\langle\hat{\mathcal L}\rangle$ for the $|B_+\rangle$ state for increasing values of $U$. 
The result shows the reduction of loop current and the breakdown of the effective theory with higher interaction values. 

\emph{Conclusions.} In this work, we have introduced vortex bound states in dimerized lattices with $\pi$-flux.
We have shown chiral properties of these states and we have investigated viable strategies to prepare them in optical lattices. 
We have also developed a scheme to measure particle currents in the presence of finite interactions of arbitrary strength.
These results are suitable to study properties of quantum states that break time-reversal symmetry, and offer an opportunity to prepare many-body chiral gapped insulating phases, for example \cite{Piraud2015, DiLiberto2023}.
Recent experimental developments in circuit-QED systems have demonstrated the control on engineering synthetic magnetic flux for interacting photons \cite{Roushan2017}, including $\pi$-flux  \cite{Houck2023}. 
Many-body gapped phases, including topological ones \cite{Wang2024Jan}, have been stabilized via non-markovian baths \cite{Biella2017Aug, Lebreuilly2017Sep,Ma2019Feb,Saxberg2022Dec}. 
Motivated by these advances, it would thus be of interest to explore the preparation of chiral photon pairs or strongly-correlated many-body chiral states in dissipative settings \cite{Tecer2024}.
Finally, as the underlying BBH model \cite{Benalcazar2017Jul} has non trivial topology, an open question concerns the consequences in the two-body spectrum, in analogy with other two-particle analyses of topological models.

\emph{Acknowledgements.}
We thank M. Aidelsburger for a feedback on the experimental aspects relevant to this manuscript. We thank I. Bloch, O. Gamayun, N. Goldman, M. Gorlach for helpful discussions.
ED results were obtained using the library QuSpin \cite{QuSpin2019}.
A.S. acknowledges support from the Priority 2030 Federal Academic Leadership Program during his work at ITMO.
M.D.L. acknowledges support from the Italian Ministry of University and Research via the Rita Levi-Montalcini program, and from Horizon Europe programme HORIZON-CL4-2022-QUANTUM-02-SGA via the project 101113690 (PASQuanS2.1) and the QuantERA project T-NiSQ.

\bibliography{article/ref}

\clearpage

\section{Adiabatic preparation}

\begin{figure}[t]
\includegraphics[width = 0.49\textwidth]{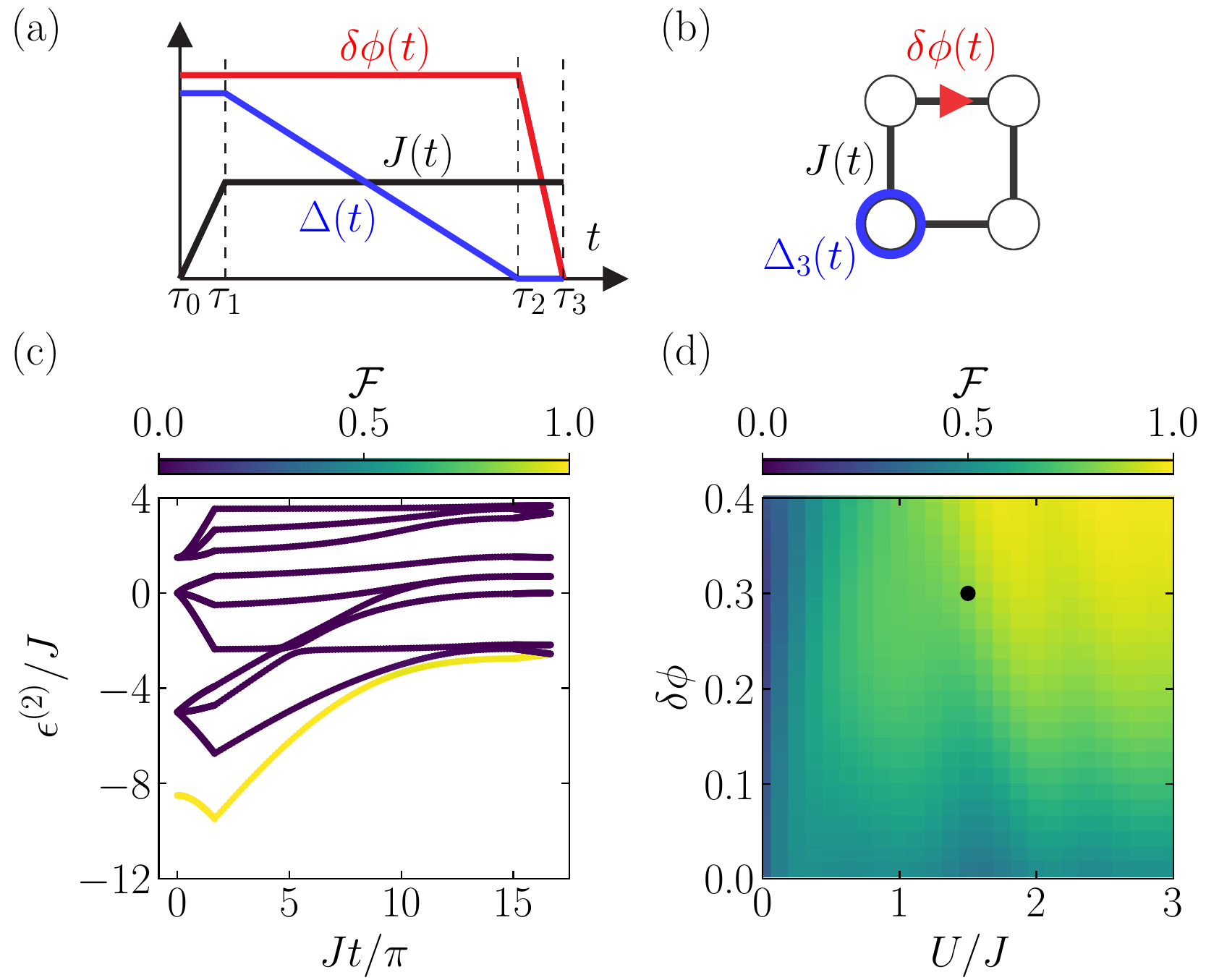}
\caption{\label{fig:prep}
Adiabatic state preparation.
(a) Sketch of the control sequence for the parameters involved in the state preparation protocol.
(b) Representation of the parameters used in (a) for a single plaquette geometry.
(c) Fidelity for $U/J=1.5$ with the instantaneous spectrum eigenstates during the time protocol showing a high ground state fidelity reached for initial states $(\cra{b}{3})^2\ket{0}/\sqrt 2$. 
(d) Fidelity maps at final time $J\tau_3 = 50\pi/3$ with parameters $\tau_3-\tau_2=\tau_1=0.1\tau_3$ and $\Delta/J = 5$ for the initial state in (c).
The black dot corresponds to the parameters $\delta\phi = 0.3$ and $U/J = 1.5$ used in panel (c).
}
\end{figure}

Here we discuss the adiabatic state preparation, which requires the control over two parameters, namely the plaquette flux and onsite energies. 
Controlling the flux allows us to split the degeneracy between chiral modes and to favour adiabaticity.
For simplicity, let us consider the Hamiltonian of a single plaquette
\begin{eqnarray}
    \hat{H}_{\text{pl}} (t) &=& \hat{H}_J + \hat{H}_U + \hat{H}_\Delta \,,
    \label{eq:Ham_single}
\end{eqnarray} 
where we dropped the plaquette index, and the various terms correspond to
\begin{eqnarray}
    \hat{H}_J &=& - J(t) (e^{-i\phi(t)}\cra{b}{1}\ana{b}{2} + \cra{b}{4}\ana{b}{2} +\cra{b}{3}\ana{b}{4}+\cra{b}{1}\ana{b}{3} + \mathrm{H.c.})    \nonumber \\
    \hat{H}_U &=&  \dfrac{U}{2} \sum_{\sigma=1}^4  \hat{n}_\sigma(\hat{n}_\sigma-1)\\
    \hat{H}_\Delta &=& -\Delta_3(t) \hat{n}_3-\Delta_4(t) \hat{n}_4\,.
\end{eqnarray}
We introduced time-dependent control parameters $\Delta_{3,4}(t)$, $\phi(t)$ and $J(t)$ and the control sequence is shown in Fig.~\ref{fig:prep}(a). 
The preparation sequence requires to start from a product state that is an eigenstate of $\hat H_\text{pl}$. 
To satisfy this requirement, we assume tunneling to be negligible at $t=\tau_0$ (e.g. by using a deep optical potential), $J(\tau_0)=0$, and we focus on two possible initial choices: $(i)$ a doublon $|\psi(\tau_0) \rangle = (\cra{b}{3})^2\ket{0}/\sqrt 2$ and $(ii)$ a product state of single occupied sites $|\psi(\tau_0) \rangle = \cra{b}{3}\cra{b}{4}\ket{0}$.

During step I of the protocol ($\tau_0<t<\tau_1$), we keep the flux detuning finite and constant, $\phi(t) = \pi + \delta\phi$ and we ramp up the tunneling coefficients $J(t) = J\times(t-\tau_0)/(\tau_1-\tau_0)$.
A strong onsite shift ($\hat H_\Delta$) is used to keep the atoms strongly localized. 
In particular, for case $(i)$ we take $\Delta_3(t) = \Delta$ and $\Delta_4=0$, while for case $(ii)$ we take $\Delta_3(t) = \Delta_4(t) = \Delta$. 
The localization is guaranteed if $\Delta \gg J$.
During step II of the protocol ($\tau_1<t<\tau_2$), we ramp down the onsite shift $\Delta(t) = \Delta \times (1- (t-\tau_1)/(\tau_2-\tau_1))$, allowing the particles to distribute uniformly across the entire plaquette.
The finite flux detuning allows us to connect with one of the two chirality sectors.
Finally, during step III of the protocol we turn off the flux detuning $\delta \phi(t) = \delta\phi \times  (1- (t-\tau_2)/(\tau_3-\tau_2))$ in order to reach the $\pi$-flux limit.
 
The protocol sequence is summarized in Fig.~\ref{fig:prep}(a)-(b). 
The two-particle spectra and the corresponding fidelity $\mathcal F$ of each state with the instantaneous state are shown for the preparation of $|B_+ \rangle $ in Fig.~\ref{fig:prep}(c) 
as a function of time for a sufficiently slow ramp (total time $J\tau_3 = 50$). 
The final ground state fidelity is larger when starting with a doublon state ($(i)$ $\mathcal{F}\approx 0.977$) and drops for 
an initial state of two single-occupied sites ($(ii)$ $\mathcal{F}\approx 0.951$).
Notice that quickly turning off the flux detuning in step III (here $\tau_3-\tau_2 = \tau_3/10$) does not affect the fidelity substantially.
The initial value of the detuning $\delta \phi$ is instead crucial, and we see from Fig.~\ref{fig:prep}(d)
that we need a sufficiently large detuning $\delta\phi$ to guarantee a high fidelity at the end of the preparation process. 

Dynamical tuning of the phase $\delta\phi(t)$ can be achieved by tuning the laser beams that are used to imprint the Peierls phases. 
The phase typically depends on the relative orientation of the laser beams with respect to the main lattice axis. 
This can be controlled e.g. by projecting the beams through a lens and having them interfere in the atomic plane. 
This results in a running-wave interference pattern with tunable angle and periodicity. 
However, when changing the phase dynamically, one has to carefully avoid any heating resonance to higher bands of the lattice potential.

\clearpage 
\onecolumngrid
\section*{Supplemental Material} 
\twocolumngrid
\setcounter{section}{0} 
\setcounter{equation}{0} 
\setcounter{figure}{0} 
\renewcommand{\thesection}{S\arabic{section}} 
\renewcommand{\theequation}{S\arabic{equation}} 
\renewcommand{\thefigure}{S\arabic{figure}} 
\section{Model}

The model discussed in this work is inspired by the construction in Ref.~\cite{DiLiberto2023} and the BBH model introduced in the context of higher-order topological insulators~\cite{Benalcazar2017Jul}. 
More specifically, we consider two bosonic particles in a one-dimensional dimerized ladder threaded by uniform flux, as shown in Fig.~\ref{fig:model}(a) and described by a Bose-Hubbard Hamiltonian
\begin{widetext}
\begin{align}
\hat{H} &= \sum_{m} \left( J_{\sigma,\sigma'} \cra{b}{m,\sigma}\ana{b}{m,\sigma'} + \textrm{H.c.} \right) +
\sum_{m} \left( J_{\sigma,\sigma'}' \cra{b}{m+1,\sigma}\ana{b}{m,\sigma'} + \textrm{H.c.} \right) 
+\dfrac{U}{2}\sum_{m,\sigma} \hat{n}_{m,\sigma}(\hat{n}_{m, \sigma}-1) \nonumber \\
&\equiv \hat H_J + \hat H_{J'} + \hat H_U \,,
\label{Seq:Ham}
\end{align}
\end{widetext}
where $\cra{b}{m,\sigma}$ and $\ana{b}{m, \sigma}$ are bosonic creation and annihilation operators in unit cell $m$ and sublattice $\sigma$, with $\sigma=1,\dots,4$. 
The non-vanishing hopping coefficients are defined as $J_{1,3}=J_{2,4}=J_{3,4}\equiv -J$, $J_{1,2}\equiv-J e^{-i\phi}$, $J'_{1,2}\equiv-J'e^{i\phi}$, $J'_{3,4}\equiv -J'$ and are pictorially indicated in Fig.~\ref{fig:model}(a).

We focus on the regime near $\pi$-flux, namely $\delta \phi \equiv \phi-\pi \ll 1$, such that the four single-particle eigenstates of $\hat H_J$ are grouped in pairs and are separated by a gap $\Delta E_\text{gap}\approx 2\sqrt{2}J$.  
Similarly to the discussion in Ref.~\cite{DiLiberto2023}, the energy-scale separation provided by the gap allows us to consider the lowest two eigenmodes $|d_{m,\pm} \rangle$ to construct a projected low-energy theory when $J'$ and $U$ are small as compared to $\Delta E_\text{gap}$.
The single particle modes display the real-space structure $|d_{m,\pm} \rangle = \sum_{\sigma=1}^4 \alpha_{\pm,\sigma} |b_{m,\sigma} \rangle$, where $\alpha_\pm = (u_\pm,v_\pm,1,u_\pm^*)/2$ with $u_\pm = e^{i(\pi\pm\pi-\phi)/4}$ and $v_\pm = e^{i(\pi\pm\pi+\phi)/2}$. 
For $\phi=\pi$, the two eigenstates are degenerate in energy and are time-reversal partners. 
Each mode breaks time-reversal symmetry and thus carries the notion of chirality. 
We quantify this physical property in real space by defining the plaquette bond currents
$\hat j^{(m)}_{\sigma,\sigma'} = i (J_{\sigma,\sigma'} \hat b^\dagger_{m,\sigma} \hat b^{}_{m,\sigma'}- \text{H.c.})$, which we translate into a loop-current operator $\mathcal{\hat L}_m = \sum_{\sigma,\sigma'}' \hat j_{\sigma,\sigma'}^{(m)}$, where $\sum'$ indicates that we have chosen to order the bond currents in a clockwise loop. 
More explicitly, the plaquette loop current operator reads
\begin{multline}
    \label{eq:curr_pl}
    \mathcal{\hat{L}}_m = i J (\cra{b}{m,4}\ana{b}{m,2}+\cra{b}{m,3}\ana{b}{m,4}+\cra{b}{m,1}\ana{b}{m,3}\\ + e^{i\phi} \cra{b}{m,2}\ana{b}{m,1}-\mathrm{H.c.})\:.
\end{multline}
Within the projected theory and for $\phi=\pi$, one can show that the loop current is related to the relative occupation of the two chiral modes, i.e. the chiral population difference $\hat{L}_{z,m} \equiv \cra{d}{m,+}\ana{d}{m,+} - \cra{d}{m,-}\ana{d}{m,-}$, via the relation $\hat{\mathcal{L}} \approx \sqrt{2}J\hat{L}_{z,m}$. 
In this sense, the two modes carry a unit of loop current with opposite signs and play a similar role as the $p_\pm = p_x\pm i p_y$ orbitals in $p$-band models~\cite{Wu2015Jun,Sala2015Mar,DiLiberto2023}. 
For this reason, we will refer to $\hat{L}_{z,m}$ as the \emph{angular momentum} operator.
However, the proportionality between the loop current operator and the angular momentum operator is not valid for $\phi\neq\pi$ or beyond the validity of the effective theory, and deviations take place, as we will show below.

Under the assumption $\delta \phi\ll 1$ and $J', U \ll J$, the Hamiltonian Eq.~\eqref{Seq:Ham} can thus be projected on the manifold spanned by the single-particle chiral modes $| d_{m,\pm} \rangle$ and reads
\begin{eqnarray}
   \hat{H}_\text{eff}&=&  
   \sum_{m,\alpha, \beta} \left( J_{\alpha,\beta}(\phi) \cra{d}{m+1,\alpha}\ana{d}{m,\beta} +\mathrm{H.c.} \right) \nonumber \\&& 
   +\sum_{m,\alpha} \mu_{\alpha}(\phi) \cra{d}{m,\alpha}\ana{d}{m,\alpha}\nonumber\\&&
    + U\sum_m\left(\dfrac{3\hat{n}_m^2}{16} - \dfrac{\hat{L}_{z,m}^2}{16}  - \dfrac{\hat{n}_m}{8} \right)\,,    
   \label{Seq:Ham_proj}
\end{eqnarray}
where $\alpha,\beta\in\left\{+,-\right\}$ label the two lowest eigenmodes and we introduced the projected particle number operator $\hat{n}_m = \cra{d}{m,+}\ana{d}{m,+} + \cra{d}{m,-}\ana{d}{m,-}$.

The first line in Eq.~\eqref{Seq:Ham_proj} determines the intercell processes with hopping amplitudes
$J_{+,+} = -J' e^{i\phi}\cos{(3\phi/4)}/2$ and
$J_{-,-} =  J'e^{i\phi}\sin{(3\phi/4)}/2$ that couple modes with the same  chirality, while the hopping amplitudes
$J_{-,+} = -J'e^{i\phi/4}(1-ie^{i3\phi/2})/4$ and
$J_{+,-} = J'e^{i\phi/4}(e^{i3\phi/2}+i)/4$ couple modes with different chirality.
In the limit $\phi\rightarrow{\pi}$ the orbitals decouple, namely $J_{\pm,\mp}\rightarrow 0 $, and $|J_{-,-}| = |J_{+,+}|$ as in Ref.~\cite{DiLiberto2023}. 
The dependence of the hopping coefficients on the flux $\phi$ is shown in Fig.~\ref{fig:model}(d).

The second line instead describes a local energy penalty for the two orbitals that provides the leading contribution to the splitting of the two orbitals and has values $\mu_+=-2J\sin{\phi/4}$ and $\mu_-=-2J\cos{\phi/4}$.

Finally, the third line describes two-particle interactions, which include a density-density term, an angular momentum term and an overall chemical potential shift. 
Differently from the single-particle terms, interactions do not display a dependence on flux. 
Furthermore, they have the same structure as the ones appearing in $p$-band collision terms of optical lattices in the tight-binding regime \cite{Liu2006}. 

\section{Two-particle bound states}
\label{sec: app_proj_disp}

\subsection{Dispersion of the projected model}
Consider the one-dimensional Hamiltonian 
\begin{eqnarray}
    \hat{H} &=& - \sum_{\alpha \in\left\{+,-\right\}} J_{\alpha} \cra{d}{m,\alpha}\ana{d}{m+1,\alpha} + \text{h.c.}
    \nonumber\\
    &&+U\sum_m\left(\dfrac{3\hat{n}_m^2}{16} - \dfrac{\hat{L}_{z,m}^2}{16}  - \dfrac{\hat{n}_m}{8} \right)\,,
\end{eqnarray}
where  $\hat{n}_m = \cra{d}{m,-}\ana{d}{m,-} + \cra{d}{m,+}\ana{d}{m,+}$ and $\hat{L}_{z,m} = \cra{d}{m,+}\ana{d}{m,+} - \cra{d}{m,-}\ana{d}{m,-}$. 
The coupling coefficient introduced here corresponds to the couplings of the modes with the same chirality $J_{\alpha}\equiv -J_{\alpha,\alpha}$ 
introduced in the main text.
Notice that here these coefficients are complex in the presence of flux. 
Moreover, we neglect $J_{\alpha,\beta}$ with $\alpha\neq\beta$ in first approximation as orbital changing processes are negligible at $\phi \approx \pi$.

Solving the Schr\"odinger equation 
$\epsilon^{(2)}\ket{\psi} = \hat{H}\ket{\psi}$ with the two-particle wavefunction ansatz in the form:
\begin{eqnarray}
    \ket{\psi} &=& \sum_{\alpha,\beta}\sum_{m,n} \theta_{m,n}^{\alpha,\beta} \cra{d}{m,\alpha} \cra{d}{n,\beta} \ket{0}\:,
\end{eqnarray}
we obtain a linear system
\begin{eqnarray}
    (\epsilon^{(2)} - U_{\alpha,\beta}) \theta_{n,n}^{\alpha,\beta} &=& -J_{\alpha} \theta_{n,n+1}^{\alpha,\beta} - J^*_{\alpha} \theta_{n,n-1}^{\alpha,\beta} \nonumber\\
    &&-J_{\beta} \theta_{n+1,n}^{\alpha,\beta} -J^*_{\beta} \theta_{n-1,n}^{\alpha,\beta} \;,
    \\
    \epsilon^{(2)} \theta_{m,n}^{\alpha,\beta} &=& -J_{\alpha} \theta_{m,n+1}^{\alpha,\beta} - J^*_{\alpha} \theta_{m,n-1}^{\alpha,\beta} \nonumber\\
    &&-J_{\beta} \theta_{m+1,n}^{\alpha,\beta} -J^*_{\beta} \theta_{m-1,n}^{\alpha,\beta}\;,
\end{eqnarray}
where $U_{\alpha,\beta}=U(2-\delta_{\alpha,\beta})/4$.

Substituting the coefficients in the form of a plane-wave ansatz 
$\theta_{m,n}^{\alpha,\beta} = C_{\alpha,\beta} e^{i k (m+n)/2}e^{i\kappa(n-m)/2}$, we obtain the equations 
\begin{eqnarray}
    (\epsilon^{(2)} - U_{\alpha,\beta})C_{\alpha,\beta}  &=& -C_{\alpha,\beta}e^{i\kappa/2}((J_{\alpha} +J_{\beta}) e^{i k /2}\nonumber\\
    &&  + (J^*_{\alpha}+J^*_{\beta}) e^{-i k /2})\,,
    \\
    \epsilon^{(2)} C_{\alpha,\beta} &=& -C_{\alpha,\beta}(e^{i k/2}(J_{\alpha} e^{i\kappa/2}+J_{\beta} e^{-i\kappa/2})\nonumber\\
    &&  + e^{-i k /2}(J^*_{\alpha} e^{-i\kappa/2}+J^*_{\beta} e^{i\kappa/2}))\,.
\end{eqnarray}
Finally, we denote $z_\alpha(k) = J_{\alpha} e^{i k/2}$  and introduce\break $y_k = 2\Re{\left[z_{\alpha}(k) + z_{\beta}(k)\right] }$ and $x_k = y_k^{-1}(z_\alpha(k)+z_\beta^*(k))$ that transform the system of equations into
\begin{eqnarray}
    (\epsilon^{(2)}(k) - U_{\alpha,\beta})  &=& -e^{i\kappa/2} y_k \;,
    \\
    \epsilon^{(2)}(k)  &=& -(x_k y_k e^{i\kappa/2}+ x_k^*y_k e^{-i\kappa/2})\,.
\end{eqnarray}
The set of equations has three solutions 
\begin{equation}
    \epsilon(k) =\dfrac{1-x_k}{2-x_k}U_{\alpha,\beta} + \sgn{(U_{\alpha,\beta})}\sqrt{\dfrac{U_{\alpha,\beta}^2}{(2-x_k)^2}+\dfrac{4x_k^*y_k^2}{2-x_k}}\:,
\end{equation}
depending on the choices of $\alpha$ and $\beta$ and it is plotted in the main text, see Fig.~\ref{fig:model}.

\subsection{\label{sec:eff} Perturbative corrections to the projected model}%

The analysis performed so far is based on the projected theory description. 
However, the high-energy plaquette modes that we have neglected will modify this picture. 
In view of the experimental impact of these effects, it is crucial to investigate the consequence of strong interactions. 
Too weak interactions would indeed hinder the possibility to practically observe chiral effects of bound pairs, whereas too strong interactions would break the projected orbital picture. 
Here, we analyse the chiral bound pairs at the single plaquette level within perturbation theory for increasing values of $U/J$ and $\phi = \pi$.
In the previous section, we show the two-particle spectrum for the full lattice system beyond the projected model limit.
There we observe the appearence of other stable bound states, originating from the contribution of the plaquette upper energy modes or more exotic ones. 

\begin{figure}[!t]
\includegraphics[width = \columnwidth]{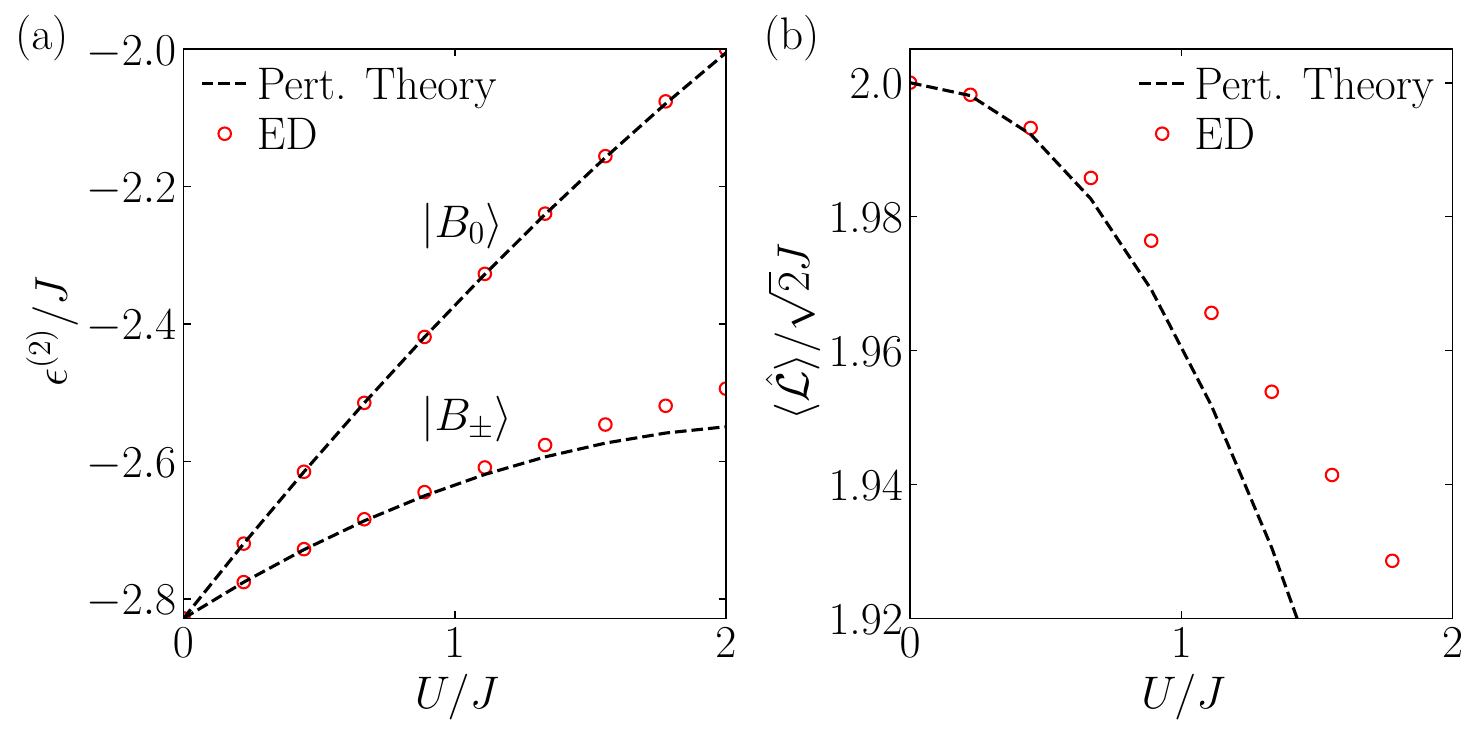}
\caption{\label{fig:eff}
Perturbative effects on the single-plaquette. (a) Single-plaquette two-particle energies $\epsilon^{(2)}$ for increasing values of interactions $U$. (b) Loop current expectation value of the $B_+$ state for increasing values of $U$.  In all panels, second-order perturbative results (dashed) are compared with the exact diagonalization results (circles).}
\end{figure}

A generic plaquette two-particle state can be expanded as $\ket{B} = \sum_n (U/J)^n \ket{B^{(n)}}$ up to second order. 
$\ket{B^{(0)}}$ indicates the non-interacting component, $\ket{B^{(1)}}$ is the perturbative correction originating from the projected theory contribution that we have already discussed and $\ket{B^{(2)}}$ takes into account virtual processes related to the high-energy single-particle eigenmodes of the plaquette.
In the case of connected plaquettes, there can also be other virtual processes to the high-energy modes.
However, we limit our analysis to the regime $J' \ll U, J$, which corresponds to bound pairs that are strongly localised on a single plaquette.
For the bound pairs energies of the states $B_\pm$ and $B_0$, we find 
\begin{eqnarray}
    \label{eq:corr1}
    \epsilon^{(2)}_\pm  &=& -2\sqrt{2}J + \dfrac{U}{4} - \dfrac{5U^2}{64\sqrt{2}J}\,,\\
    \epsilon^{(2)}_0 &=& -2\sqrt{2}J + \dfrac{U}{2} - \dfrac{U^2}{16\sqrt{2}J}\,,
\end{eqnarray}
which we plot in Fig.~\ref{fig:eff}(a) for $U<2J$. 
The Figure shows that second-order perturbative corrections of the energies are valid up to large values of $U/J$, and that the two types of bound states ($B_\pm$ and $B_0$) are energetically well resolved.

In order to verify whether the orbital order survives the strong interaction, we determine the loop current value of the bound states, which reads
\begin{eqnarray}
    \bra{B_\pm} \mathcal{\hat{L}} \ket{B_\pm}  &=& \pm \left(2\sqrt{2} - \dfrac{5U^2}{64\sqrt{2}J^2}\right)\\
    \label{eq:corr4}\bra{B_0} \mathcal{\hat{L}} \ket{B_0}  &=& 0
\end{eqnarray}
A comparison with the numerically exact solution, Fig.~\ref{fig:eff}(b) shows that perturbative corrections for the loop current are more severe for this range of parameters. 
A positive conclusion that can be drawn from the numerically exact results is that the loop current is actually less affected by the higher modes contribution when increasing $U/J$.
Higher orders in perturbation theory compensates the drop from the second-order contribution and keep the loop current of the state closer to its maximum value, thus providing a potentially stronger signal for experimental detection.

\subsection{Beyond projected model}

\begin{figure}
    \centering
    \includegraphics[width=\linewidth]{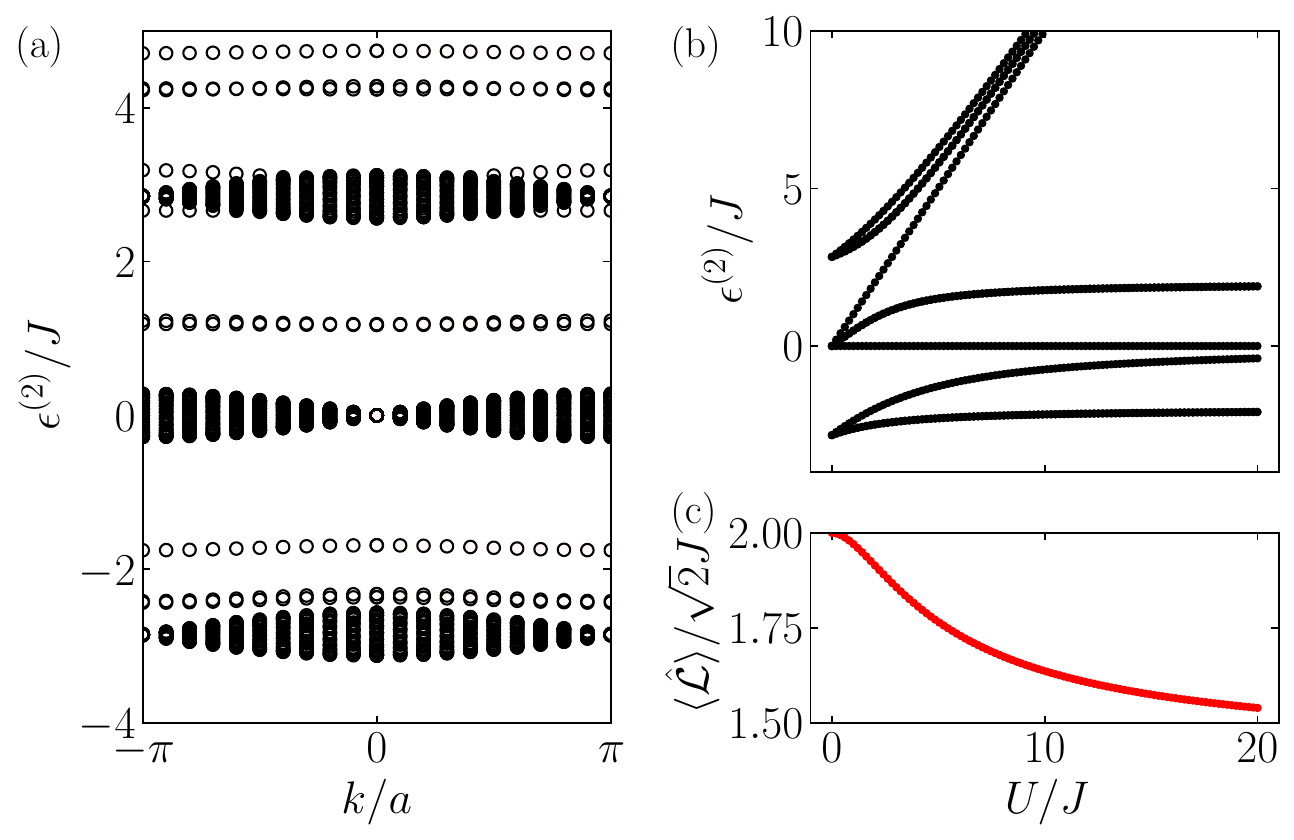}
    \caption{
    \label{fig:full_2body}
    (a) Two body spectrum for a finite array of $N=20$ plaquettes with $J'/J=0.2$, $U/J=3$ and $\delta\phi=10^{-3}$. (b) Single-plaquette spectrum for increasing value of interactions $U$. (c) Loop current of a vortex two-body state, corresponding to the lowest eigenvalue in (b), as a function of interactions up to regimes where projected theory is not 
 valid anymore. A small flux detuning $\delta\phi = 10^{-3}$ was used to split the degeneracy.}
    \label{fig:app1}
\end{figure}

While most of the results discussed in this work focus on the low-energy two-particle spectrum and the connection to the projected theory description, a representative full two-body spectrum is shown in Fig.~\ref{fig:full_2body}(a) for $U/J=3$ and $J'/J=0.2$.
We identify three scattering continua, originating from the single-particle bands shown in Fig.~\ref{fig:full_2body}(a).
In analogy with the notation $|d_{\pm}\rangle$ used for the single-particle negative-energy modes, we indicate the positive energy ones as $|u_{\pm}\rangle$.
The negative energy continuum displays the properties that we have analysed so far within the projected theory based on the lowest two single-particle (degenerate) bands and the bound states $B_\pm$ and $B_0$.
The zero energy continuum is four-fold degenerate, as it originates from single-particle states taken from the two lower and upper bands.
This also occurs for bound states, for example those at energy $\epsilon^{(2)}\sim J$ in Fig.~\ref{fig:full_2body}(a). 
The high-energy sector displays interesting novel features. Besides the very high-energy modes, which correspond to the analogous $B_\pm$ and $B_0$ states for the upper bands, we also notice new bands quasi-resonant with the upper scattering continuum at energy $\epsilon^{(2)}\sim 2\sqrt 2 J$.
This behaviour shares similarities with the one identified in the SSH model,  see Ref.~\cite{DiLiberto2016}, where analogous states appear and it requires a more elaborate analysis that is beyond the scope of this work.

In Fig.~\ref{fig:full_2body}(b), the single-plaquette spectrum is shown, from which one can deduce some of the features, i.e. the bound states, shown in Fig.~\ref{fig:full_2body}(a). 
We plot the results up to very large interactions, where new effective descriptions can take place. For example the very high-energy modes can be described in terms of doublons, namely doubly-occupied sites.
We anticipate that, in this regime, the model is actually effectively describing an SSH-like model of doublons (either in a ladder or a full 2D scenario) due to the flux doubling that doublons experience (shown e.g. in Ref.~\cite{Salerno2018}). 
In this case, no actual flux will be present for the pairs in their effective description, showing a completely novel scenario and possibly intriguing topological effects. 

Finally, in Fig.~\ref{fig:full_2body}(c) we follow the $|B_+\rangle$ state up to very high interaction values where its energy flattens (see Fig.~\ref{fig:full_2body}(b). 
We plot the corresponding loop current value finding that it does not drop substantially. 
This occurs even though the nature of the two-particle state is very different from the original picture based on the emergent angular momentum interaction, providing a quantization of $\langle \hat L_z \rangle \approx 2$.

\section{Current and differential density}

\label{sec:app_current_meas}

In this section, we provide an exact derivation of the expectation values of the current $\langle\hat{j}_{ab}\rangle$ and differential density $\langle\hat{n}_{ab}\rangle = \langle\hat{n}_a - \hat{n}_b\rangle$ for two coupled sites, described by the \emph{dimer} Hamiltonian
\begin{eqnarray}
    \hat{H}_{ab} &=& -(J_{ab}\cra{b}{a}\ana{b}{b}+J_{ab}^*\cra{b}{b}\ana{b}{a}) \nonumber\\
    &&+ \dfrac{U}{2} (\hat{n}_a(\hat{n}_a-1)+\hat{n}_b(\hat{n}_b-1))\,,
    \label{eq:H_dim}
\end{eqnarray}
with complex hopping $J_{ab}=Je^{i\phi}$. 
A gauge transformation can make the hopping coefficients real, but we will keep the general treatment with complex ones.
For completeness, the current operator is here defined as $\hat{j}_{ab} = i(J_{ab}\cra{b}{a}\ana{b}{b}-J_{ab}^*\cra{b}{b}\ana{b}{a})$.

\subsection{Dimer symmetry analysis}

The dimer Hamiltonian~\eqref{eq:H_dim} commutes with the parity operator $\hat P$ that exchanges sites $a$ and $b$:
\begin{eqnarray}
    \hat{P} \ana{b}{a}\hat{P}^\dagger &=& \ana{b}{b}e^{i\phi}\,,\\
    \hat{P} \ana{b}{a}\hat{P}^\dagger &=& \ana{b}{b}e^{-i\phi}\,,
\end{eqnarray}
and 
\begin{eqnarray}
    \hat{P}\ket{\psi_{\pm}} &=& \pm\ket{\psi_{\pm}}\:,
\end{eqnarray}
where $|\psi_\pm\>$ are eigenstates of the parity operator with given parity. 
While the Hamiltonian commutes with the parity, $\left[\hat{H}_{ab},\hat{P}\right] = 0$, the differential density anticommutes $\left\{\hat{n}_{ab},\hat{P}\right\} =0$.
This property will be useful later.

\subsection{Single-particle}

Before addressing the interacting case, we first revisit the single-particle (or noninteracting) result~\cite{Atala2014Aug}.
Let us denote the two single-particle dimer eigenstates as $\ket{u_{1,2}} = (1,\pm e^{-i\phi})^T/\sqrt{2}$ with energy $\eps_{1,2} = \mp J$, which have opposite parity. 
The generic time-dependent solution reads
\begin{eqnarray}
    \ket{\psi} 
    &=&c_1 \ket{u_1} e^{-i\eps_1 t} + c_2 \ket{u_2} e^{-i\eps_2 t}\,,
\end{eqnarray}
with coefficients $c_{1} = \<u_{1}|\psi_0\>$, $c_{2} = \<u_{2}|\psi_0\>$ and $|\psi_0\>$ the state at $t=0$.
We thus arrive at
\begin{eqnarray}
    \langle\hat{n}_{ab}\rangle(t) &=&  c^*_1 c_2 e^{-2iJ t}  + c^*_2 c_1 e^{2iJ t}\:,
    \\
    \langle\hat{j}_{ab}\rangle(t) &=& iJ(c_2^*c_1 e^{2iJ t}+ c_1^*c_2 e^{-2iJ t}) \,,
\end{eqnarray}
 which can be recast in the form
\begin{eqnarray}
    \langle\hat{n}_{ab}\rangle(t) &=& \mathcal{N}_0^{(1)}\cos(2Jt) + \dfrac{\mathcal{J}_0^{(1)}}{J}\sin(2Jt)\:,
    \label{eq:Ssp1}\\
    \langle\hat{j}_{ab}\rangle(t) &=&  -J \mathcal{N}_0^{(1)}\sin(2Jt) + \mathcal{J}_0^{(1)}\cos(2Jt) \:,
    \label{eq:Ssp2}
\end{eqnarray}
where $\mathcal{N}_0^{(1)}$ and $\mathcal{J}_0^{(1)}$ denote the single-particle differential density and current at $t=0$, respectively.
Note that the equation for the current can be deduced from the continuity equation, which justifies the fact that the equations display only two independent parameters.
Indeed, from $2\langle\hat{n}_{a}\rangle = N + \langle\hat{n}_{ab}\rangle$, where $N=1$ here, and the continuity equation
$\partial_t \langle\hat{n}_{a}\rangle=\langle\hat{j}_{ab}\rangle$, we immediately obtain Eq.~\eqref{eq:Ssp2} from Eq.~\eqref{eq:Ssp1}.
Let us denote $\mathcal N_1^{(1)} \equiv \langle\hat{n}_{ab}\rangle(\tau_1)$, with $\tau_1=\pi/4J$.
From Eq.~\eqref{eq:Ssp1}, we immediately find that $\mathcal J_0^{(1)} = J \mathcal N_1^{(1)}$.
This result tells us that we can deduce the value of the current at time $t=0$ by measuring the differential density at time $\tau_1$. 
Notice that one needs also to know the value of the hopping parameter.

\subsection{Two-particle}

The two-particle Hilbert space for the Hamiltonian \eqref{eq:H_dim} is spanned by the following states:
\begin{equation}
\hat b^\dagger_a \hat b^\dagger_a |0\>/\sqrt 2\,,\quad \hat b^\dagger_b \hat b^\dagger_b |0\>/\sqrt 2\,,\quad \hat b^\dagger_a \hat b^\dagger_b |0\>\,.
\end{equation}
Based on the parity symmetry $\hat P$, there is one \emph{dark} state of negative parity
\begin{equation}
    |d,-\> \equiv \frac{e^{i\phi}\hat b^\dagger_a \hat b^\dagger_a - e^{-i\phi}\hat b^\dagger_b \hat b^\dagger_b }{\sqrt 2} |0\>\,,
\end{equation}
and one can construct the following two states of positive parity
\begin{align}
    |s,+\> &\equiv \frac{e^{i\phi}\hat b^\dagger_a \hat b^\dagger_a + e^{-i\phi}\hat b^\dagger_b \hat b^\dagger_b }{\sqrt 2} |0\>\,,\\
    |p,+\> &\equiv \hat b^\dagger_a \hat b^\dagger_b |0\>\,.
\end{align}

The dark state, $|d,-\>$, being the only one with negative parity is also an eigenstate of the Hamiltonian, with energy $\eps_d = U$.
The states of positive parity define a two-dimensional subspace that can be diagonalized yielding the eigenvalues $\eps_{1,2} = U/2 \pm \sqrt{U^2/4 +4J^2}$ and the positive parity eigenstates of the Hamiltonian that we denote as $|1,+\>$ and $|2,+\>$.

Since the parity operator anticommutes with the differential density, $\left\{\hat{n}_{ab},\hat{P}\right\} =0$, the only nonvanishing matrix elements of $\langle\hat{n}_{ab}\rangle$ are those among states of opposite parity.
We can thus very generally write the time evolution of the differential density as 
\begin{equation}
\label{eq:nab2}
\langle\hat{n}_{ab}\rangle(t) = A \cos(\omega_1 t) + B  \sin(\omega_1 t) + C  \cos(\omega_2 t) + D  \sin(\omega_2 t)\,,
\end{equation}
where $\omega_{1,2}\equiv \varepsilon_d-\varepsilon_{1,2}$ and no frequency $\varepsilon_1-\varepsilon_2$ appears due to the parity argument given above.
We thus see that the dynamics is determined by only four parameters or amplitudes.
One operational possibility for experiments is thus to choose four arbitrary times and numerically solve the corresponding linear system.
However, here we show a particularly simple analytical solution that can be obtained by taking specific time values.

Let us define the frequency $\omega = (\omega_2-\omega_1)/2$ and let us define the four measurement times $\tau_n = n\pi/2\omega$, with $n=0,\dots,3$.
Let us call the corresponding differential measured densities as $\mathcal{N}_{n}^{(2)}$.
From \eqref{eq:nab2}, we can thus write the linear system
\begin{equation}
\label{eq:linsys}
    \left(\begin{array}{c}
     \mathcal{N}_0^{(2)} \\
     \mathcal{N}_1^{(2)} \\
     \mathcal{N}_2^{(2)} \\
     \mathcal{N}_3^{(2)}
    \end{array}\right) = \bold M
        \left(\begin{array}{c}
     A \\ [.1cm] B \\ [.1cm] C \\ [.1cm] D
    \end{array}\right) \,,
\end{equation}
where the matrix $\bold M$ reads
\begin{equation}
    \bold M \equiv 
    \left(\begin{array}{cccc}
    1 & 0 & 1 & 0 \\
    \cos{\left(\frac{\pi \omega_1}{2\omega}\right)} & 
    \sin{\left(\frac{\pi \omega_1}{2\omega}\right)} &
    \cos{\left(\frac{\pi \omega_2}{2\omega}\right)} &
    \sin{\left(\frac{\pi \omega_2}{2\omega}\right)} \\ [.1cm]
    \cos{\left(\frac{\pi \omega_1}{\omega}\right)} & 
    \sin{\left(\frac{\pi \omega_1}{\omega}\right)} &
    \cos{\left(\frac{\pi \omega_2}{\omega}\right)} &
    \sin{\left(\frac{\pi \omega_2}{\omega}\right)} \\ [.1cm]
    \cos{\left(\frac{3\pi \omega_1}{2\omega}\right)} & 
    \sin{\left(\frac{3\pi \omega_1}{2\omega}\right)} &
    \cos{\left(\frac{3\pi \omega_2}{2\omega}\right)} &
    \sin{\left(\frac{3\pi \omega_2}{2\omega}\right)}
    \end{array}\right). \nonumber
\end{equation}
This matrix can be analytically inverted (solution not shown here), and one can therefore determine the parameters $A$, $B$, $C$ and $D$.
From the continuity equation, we see that the initial current satisfies $2 \langle\hat{j}_{ab}\rangle(0) \equiv 2\mathcal J_0^{(2)} = \partial_t\langle\hat{n}_{ab}\rangle(0) = \omega_1 B + \omega_2 D$. 
Collecting all the results together, we thus find
\begin{align} 
    \mathcal{J}_{0}^{(2)} 
    =&-\csc{\left(\frac{U\pi}{2\omega}\right)}\left[
    \dfrac{U}{4}\mathcal{N}_{0}^{(2)}
    \cos{\left(\frac{U\pi}{2\omega}\right)}\right.
    +\dfrac{U}{4}\mathcal{N}_{2}^{(2)}\nonumber\\&
    -\dfrac{\omega}{2} \mathcal{N}_{1}^{(2)}\sin{\left(\frac{3U\pi}{4\omega}\right)}
    \left.-\dfrac{\omega}{2} \mathcal{N}_{3}^{(2)}\sin{\left(\frac{U\pi}{4\omega}\right)}
    \right]\,.\nonumber
\end{align}
In the limit $U\rightarrow 0$, this equation reduces to the one found in the previous section for the noninteracting (single-particle) case. 

Notice that the number of required measurements could be further reduced. 
This can be achieved by changing the measurement time for the single-particle dynamics, chosen to be $\tau=\pi/4J$, to one of those used for two-particle dynamics, namely $\tau_n=n\pi/2\omega$ provided that the differential density is large enough to minimize the signal-to-noise ratio and the corresponding associated error.
However, the formula connecting the single-particle current with the differential density introduced above, would not be valid anymore.
Instead, from Eq.~\eqref{eq:Ssp2}, one would obtain $\mathcal{J}_0^{(1)} = J(\mathcal{N}_{\tau_n}^{(1)} -\mathcal{N}_0^{(1)}\cos(2J\tau_n) )/\sin(2J\tau_n)$.

\section{State preparation using current imprint}

\label{sec:current_imprint}

This section discusses the general protocol for preparing a single-particle eigenstate of the $\pi$-flux plaquette with given chirality and uniform density.
Let us start with a particle localized at one of the four sites of the plaquette, which we denote as $m$.
To keep the algorithm general for any choice of $m\in\left[1,4\right]$ we consider the vector $\ket{\psi_0} = (x_1,x_2,x_3,x_4)^T$ as initial state, where the coefficients satisfy $x_{m'} = \delta_{mm'}$.

Consider two unitary operations over two sites $a$ and $b$ 
\begin{eqnarray}
    \hat{\mathcal U}_{ab}^X(t)&\equiv& \exp\left(-i (-J_{ab}\cra{b}{a}\ana{b}{b}^{}+\text{h.c.})t/\hbar\right)\,,\\
    \hat{\mathcal U}_{ab}^Z(t)&\equiv& \exp\left(-i \Delta (\hat{n}_a - \hat{n}_b)t/2\hbar\right)\,.
\end{eqnarray}

To target the phase pattern of the chiral state, we will follow a similar sequence of steps as presented in the main text.
For simplicity of notation we take $J_{ab}=J=\Delta=\hbar=1$.

\begin{enumerate}[label=\Roman*.]
    \item A particle, initially localized on site $m$, is transported to site $m'$ via a combined pulse $\hat{\mathcal U}_{12}^X(\pi/4)\hat{\mathcal U}_{34}^X(\pi/4)$;
    \item The relative phase is adjusted via a pulse $\hat{\mathcal U}_{12}^Z(\phi_1)\hat{\mathcal U}_{34}^Z(\phi_1)$;
    \item The particle is transferred to the empty sites by a combined pulse $\hat{\mathcal U}_{31}^X(\pi/4)\, \hat{\mathcal U}_{42}^X(\pi/4)$;
    \item The phases of the newly populated sites are shifted by $\hat{\mathcal U}_{13}^Z(\phi_2)\, \hat{\mathcal U}_{24}^Z(\phi_2)$;
    \item The phases of the \emph{diagonal sites} $1$ and $4$ are shifted by 
    $\hat{\mathcal U}_{13}^Z(\phi_3)\, \hat{\mathcal U}_{24}^Z(-\phi_3)$;
\end{enumerate}

Note, that these steps require phase shifts $\phi_1,\phi_2,\phi_3$ that depend on the choice of the initial position $m$ of the particle and the target state.

To be specific, we consider a single $\pi$-flux plaquette with hopping coefficients $J_{1,3}=J_{2,4}=J_{3,4}\equiv -J$, $J_{1,2}\equiv-J e^{i\pi}$. 
After the application of the five unitary operations discussed above, the single-particle state $\ket{\psi_0}$ is transformed into
\begin{eqnarray}
        \ket{\psi} = \dfrac{1}{2}\left(\begin{array}{c}
            (x_1 - i x_2 + ix_3 - x_4)e^{i\phi_2}e^{i\phi_3} \\
            (-ix_1+x_2-x_3+ix_4)e^{-i\phi_1}e^{i\phi_2}\\
            x_3+ix_4+i x_1 + x_2\\
            (ix_3+x_4+x_1+ix_2)e^{-i\phi_1}e^{i\phi_3}
        \end{array}\right)\,.
    \end{eqnarray}
For example, by choosing $m=3$ for the initial state, and thus $x_3 = 1$ and all other components zero, the final state results in
\begin{eqnarray}
    \ket{\psi} = \dfrac{1}{2}\left(\begin{array}{c}
            e^{i(\phi_2+\phi_3+\pi/2)} \\
            e^{i(\phi_1+\phi_2+\pi)}\\
            1\\
            e^{i(\phi_1+\phi_3+\pi/2)}
        \end{array}\right) \equiv \frac 1 2
        \left(\begin{array}{c}
            e^{i\theta_1} \\
            e^{i\theta_2}\\
            1\\
            e^{i\theta_3}
        \end{array}\right)
        \:.
\end{eqnarray}
We can thus achieve the desired final phase pattern ($\theta_{1,2,3}$) by solving the linear system
\begin{eqnarray}
\begin{cases}
    \phi_2+\phi_3+\pi/2 = \theta_1\\
    \phi_1+\phi_2+\pi = \theta_2\\
    \phi_1+\phi_3+\pi/2 = \theta_3
\end{cases}\,,
\end{eqnarray}
which has as the solution
\begin{eqnarray}
\begin{cases}
    \phi_1 = (\theta_2 + \theta_3 - \theta_1-\pi)/2\\
    \phi_2 = (\theta_2 - \theta_3 + \theta_1-\pi)/2 \\
    \phi_3 = (\theta_3 - \theta_2 + \theta_1)/2 
\end{cases}\,.
\end{eqnarray}
Now, choosing the target state $\ket{\psi} = \ket{d_+}$, whose corresponding phases $\theta_{1,2,3}$ are given by $|d_+\rangle \sim e^{i\pi/4}|b_1\rangle + e^{i3\pi/2}|b_2\rangle + |b_3\rangle + e^{-i\pi/4}|b_4\rangle$, we find that the phases of the unitary transformations are to be chosen as $\phi_1 = \pi, \phi_2 = -\pi/2, \phi_3 = \pi/4$.
With a similar sequence, one can prepare $\ket{\psi} = \ket{d_-}$ using $\phi_1 = 0, \phi_2 = -\pi/2, \phi_3 = -\pi/4$.
As $\phi_1=0$ corresponds to no pulse, we discover that we can skip step II. 
Analogously one can verify that by starting from site $m=4$ for preparing $|d_+\>$, one finds again $\phi_1=0$ and can thus skip step II. 
This sequence is the one used in the main text. 

By inspection of the sequence of pulses used to prepare $|d_+\>$ starting from $m=3$ or $m=4$, we see that they never involve actual transport or hopping across the sites 1 and 2. 
This can offer an advantage for state preparation, since flux engineering requires, for example, the application of Floquet drives. 
One can thus envisage to start the preparation protocol with the Floquet drive off and thus all hopping coefficients real and positive. 
Once the desired phase pattern corresponding to the superposition state $|d_+\>$ is prepared, the Floquet drive is turned on and the $|d_+\>$ becomes an eigenstate of the driven Hamiltonian. 
Since the chiral state phase pattern is gauge dependent, namely it depends on the Hamiltonian hopping coefficients $J_{ab}$, it is necessary to know the Floquet Hamiltonian hopping coefficients that the drive will generate in order to design the appropriate phases $\phi_{1,2,3}$.

Another possibility is that the Floquet drive is always on and the coefficients $J_{ab}$ are therefore those of the $\pi$-flux Hamiltonian.
As discussed above, we have chosen the $\pi$-flux to be represented by a negative $J_{12}$, but any gauge choice would do.
It is straightforward to verify that the five steps identified above are still valid with the same unitaries, and the resulting phases $\phi_{1,2,3}$ are unchanged if another gauge choice is used.
The only difference will be that the phase pattern of the final state will be different due to the basis change.

\section{Towards state preparation of $B_0$} 

Our protocols, adiabatic or digital ones, can be more generally employed to prepare also a $|d_-\rangle$ state. 
Instead, the preparation of a zero angular momentum two-particle state $\hat d^\dagger_+\hat d^\dagger_-|0\rangle$, which can then be connected to the $|B_0\rangle$ state, is quite nontrivial and is left to a future analysis. 
However, we suggest two possibilities that could be pursued: $i)$ using distinguishable particles, e.g. atoms in different internal states or in two distinct plaquettes, that are independently controlled, e.g. via state-dependent lattices, and then merged; $ii)$ considering the three states $|B_0\rangle$, $|B_\pm\rangle$ as a three level system. 
For realizing $ii)$, after preparing both particles in the $|B_+\rangle$ state, a single-particle resonant drive is needed to couple states with opposite chiralities (e.g. circular lattice shaking). 
This drive is thus employed to make a state transfer within a plaquette by flipping the chirality of one of the two atoms, thus leading to the transfer $|B_+\>\rightarrow |B_0\>$. 
A detuning splitting the two states $|B_\pm\rangle$, e.g. via finite flux, has to be employed to avoid populating the state $|B_-\rangle$ as in typical Raman-like schemes.


\end{document}